\documentclass[twocolumn,preprintnumbers,amsmath,amssymb,superscriptaddress,footnote]{revtex4-2}
\usepackage{graphicx} 
\usepackage[dvipdfmx]{epsfig} 
\usepackage{bm}
\usepackage{ulem}
\usepackage{amsmath,color}

\def\m@thcombine#1#2{%
  \setbox0=\hbox{$#1$}
  \setbox1=\hbox{$#2$}
  \ifdim\wd0>\wd1
    \setbox0=\hbox to\wd1{\hss\box0\hss}
  \else
    \setbox1=\hbox to\wd0{\hss\box1\hss}
  \fi
  \mathop{\vcenter{
    \offinterlineskip\box0\box1}}}
\def\lesim{\m@thcombine<\sim}
\def\gesim{\m@thcombine>\sim}
\def\vr{\mbox{\boldmath$r$}}
\def\vJ{\mbox{\boldmath$J$}}

\begin{document}
\title{Pairing core swelling effect in Pb isotopes at $N>126$}
\author{W. Horiuchi}
\email{whoriuchi@nucl.sci.hokudai.ac.jp}
\affiliation{Department of Physics,
  Hokkaido University, Sapporo 060-0810, Japan}
\author{T. Inakura}
\affiliation{Laboratory for Zero-Carbon Energy, Institute of Innovative Research, Tokyo Institute of Technology, Tokyo 152-8550, Japan}

\begin{abstract}
We revisit a sudden increase of the isotope shift of the charge radius
of Pb isotope at $N>126$ based on a Skyrme Hartree-Fock-Bogoliubov theory.
New parametrizations of the pairing 
interaction optimized for selected four Skyrme interactions
greatly improve a description of this phenomenon.
The density-dependent spin-orbit interaction is also investigated
  and further increases the charge radius.
The pairing correlations significantly change
the properties of the neutron orbits near the Fermi level
and play a vital role in pulling out the well-bound protons
in the Pb isotopes at $N>126$.
Regarding $^{208}$Pb as a ``core'' nucleus,
a novel pairing core swelling effect is proposed:
The pairing interaction reduces the radius of ``valence'' neutron orbits
by the shrinkage of diffused $1g_{9/2}$ orbit
and the mixing of sharp $0i_{11/2}$ orbit.
Simultaneously, the core nucleus swells,
leading to the sudden enhancement of the charge radius at $N>126$.
This characteristic behavior appears in the density profile
near the nuclear surface and its measurement is highly desirable.
\end{abstract}

\maketitle

\section{Introduction}

The nuclear radii are fundamental quantities
that reflect characteristics of the density profile
of atomic nuclei. 
Systematic measurements of nuclear charge radii
have been done by using electron scattering for stable nuclei
and have extended for short-lived unstable nuclei
with isotope-shift measurements.
It is known that the mass number $A$ dependence of 
the charge radius is roughly proportional to $A^{1/3}$.
For spherical nuclei, this slope changes across the magic numbers,
where the major shell changes~\cite{Angeli13}.
Recently, a sudden increase or a kink of
the charge radius at $N>28$ was found
in the charge radius measurement of Ca isotopes~\cite{Ruiz16}
and has attracted much attention.
The phenomenon was interpreted as the swelling of a $^{48}$Ca ``core''
in the interaction cross section measurement~\cite{Tanaka20}.
Afterward, this mechanism is related
to the saturation of the internal density~\cite{Horiuchi20}.
However, it has not been reached at the universal understanding
for all mass regions including heavier mass nuclei such as Pb isotopes.

The charge radius kink observed in Pb isotopes at $N>126$
is one of the most famous examples.
For describing the properties of such heavy mass nuclei,
a Skyrme-type density functional~\cite{Vautherin72} has often been used.
Reference~\cite{Reinhard94} pointed out that
a conventional Skyrme Hartree-Fock calculation
cannot reproduce the kink behavior
and introduced the isospin-dependent spin-orbit interaction.
Reference~\cite{Nakada15a} introduced density-dependent
spin-orbit (DDLS) interaction
and showed that the kink behavior was better explained by
considering both the configuration mixing and modification 
of the $1g_{9/2}$ and $0i_{11/2}$ orbits.
We remark that the Fayans effective interaction reproduces
the kink behavior of the charge radii~\cite{Fayans00} 
but the relation of those kinks to the interaction was not clear.
Recently, the charge radius kink at $N>126$ was observed
also in Hg isotopes~\cite{Goodacre21a,Goodacre21b}
as it shows the charge radius kink is a universal phenomenon
whose emergence mechanism should be clarified.

In this paper, we revisit the kink behavior of
Pb isotopes at $N>126$ based on a Hartree-Fock-Bogoliubov (HFB) 
theory with a Skyrme density functional.
To describe the charge radius kink of heavy mass regions, 
it is vital to properly treat the pairing and spin-orbit 
interactions~\cite{Reinhard94,Nakada15a}.
The purpose of this paper is twofold:
(i) We give appropriate parametrizations for the pairing and
DDLS interactions optimized for several standard Skyrme interactions,
and then (ii) analyze size properties of the Pb isotopes 
and clarify the role of the pairing correlations and DDLS interaction
in producing the charge radius kink.

The paper is organized as follows.
Section~\ref{formalism.sec} briefly describes
the HFB theory with the Skyrme-density functional. 
The pairing and DDLS interactions include some free parameters.
Section~\ref{adjpara.sec} explains
how to fix them by using experimental data.
Section~\ref{results.sec} presents our results.
Results of the isotope shift of the charge radius are
compared with experimental data in Sec.~\ref{charge.sec}.
The importance of the pairing correlations is emphasized
to reproduce the charge radius kink of the Pb isotopes.
The DDLS further enhances the charge radius
though it is not significant than the pairing correlations.
The kink at $N>126$ is induced by the pairing interaction
and is interpreted as a swelling of the ``core''.
Section~\ref{swelling.sec} describes
the mechanism of this pairing core swelling phenomenon in detail.
The influence on observables other than the charge radius is discussed
in Sec.~\ref{diff.sec} for future measurements.
The conclusion is given in Sec.~\ref{conclusion.sec}.

\section{Theoretical models}
\label{formalism.sec}

The density functional theory (DFT) is one of the standard approaches 
to investigate nuclear properties in a systematic way. 
The DFT with the Skyrme functional describes the ground-state energy 
as a functional of three one-body densities, i.e.,
nucleon $\rho_q(\vr)$, kinetic $\tau_q(\vr)$, 
and spin-orbit $\vJ_q(\vr)$ densities, where $q$ denotes
neutron ($n$) or proton ($n$).
In the HFB formalism with spherical symmetry,
the Hamiltonian is reduced to the mean field $h$ (particle-hole channel) and 
the pairing field $\Delta$ (particle-particle channel) as
\begin{align}
\left(\begin{array}{cc}
\left( h - \lambda \right) & \Delta \\
-\Delta^\ast & -\left( h - \lambda \right)^\ast
\end{array}\right)
\left(\begin{array}{c}
U_k(r) \\ V_k(r)
\end{array}\right)
=E_k
\left(\begin{array}{c}
U_k(r) \\ V_k(r)
\end{array}\right) \,,
\end{align}
where $U_k$ and $V_k$ are the two components 
of the single quasi-particle radial wave functions and
$E_k$ is their energy, and $\lambda$ 
is the chemical potential. We use $k=nlj$  as a short-handed notation for
the quantum numbers of the system, where $n$, $l$, and $j$
are the number of nodes, 
orbital angular momentum, and total angular momentum, respectively.

Calculations are performed for Pb isotopes by using the HFB 
solver HFBRAD~\cite{Bennaceur05}, which solves the Skyrme-HFB equations
in the coordinate representation with spherical symmetry.
The model space is set to the radius $R_\mathrm{box}=25$ fm
with mesh spacing 0.1 fm. 
The maximum values for the angular momentum of the quasi-particle orbits
are set to be $j_\mathrm{max}= (55/2) \hbar$
for neutron and $(45/2) \hbar$ for proton.
We choose four standard Skyrme functionals in the particle-hole channels: 
SkM$^\ast$~\cite{SkMs},  SLy4 \cite{SLy4}, SV-min \cite{SVmin},
and UNEDF1 \cite{UNEDF1}.
The SkM$^\ast$ and SLy4 interactions have been widely used
for nuclear structure calculations, and SV-min and UNEDF1 are recently
designed as sophisticated versions of Skyrme parameter sets.
For the particle-particle channel, we employ the mixed-type pairing 
\begin{align}
V_{\rm pair}(\vr,\vr^\prime) = V_0 \left( 1 - \frac{1}{2}\frac{\rho(\vr)}{\rho_0}\right) \delta(\vr-\vr^\prime),
\label{mixed.eq}
\end{align}
where $\rho_0 = 0.16 \,\mbox{fm}^{-3}$ and $V_0$ is the pairing strength
that will be adjusted later.
  All the single-quasi-particle states
  below a cut-off energy $E_\mathrm{cut}=60$ MeV
  are considered for all the interactions,
  while a state-dependent energy cut-off
was conventionally used for SV-min~\cite{SVmin}.
Therefore, our results labeled with SV-min are somewhat different from
those with the original SV-min interaction.

References~\cite{Nakada15a,Nakada15b} suggest the need for 
the density-dependent spin-orbit (DDLS) interaction
to explain the kink behavior in the isotope shift of the charge radius
of Pb isotopes. Here we consider adopting it to the Skyrme functionals.
The spin-orbit energy functional of the Skyrme interaction is explicitly
written by \cite{Vautherin72}
\begin{align}
  E_{\rm LS}&=-\int d\bm{r}\, W_{\rm LS}\notag\\
  &\times \left\{
  \rho\nabla\cdot\bm{J}+\sum_q\rho_q\nabla\cdot\bm{J}_q
  -\bm{J}\cdot\nabla\rho-\sum_q\bm{J}_q\cdot\nabla\rho_q\right\} \,,
\end{align}
where $\rho = \rho_n + \rho_p$ and $\bm{J}=\bm{J}_n+\bm{J}_p$ with
\begin{align}
W_{\rm LS}=\frac{1}{4}W_0 \,.
\end{align}
This $W_{\rm LS}$ term is simply replaced by~\cite{Nakada15a,Nakada15b}
\begin{align}
  W_{\rm DDLS}=f_{ls}\frac{W_0}{4}+g_{ls}D[\rho(\bm{r})],
\label{DDLSimplementation}
\end{align}
where $f_{ls}$ and $g_{ls}$ are free parameters to be determined later.
A functional form of the density-dependent term
$D[\rho(\bm{r})]$ is taken as
\begin{align}
  D[\rho(\bm{r})]=w_1\frac{\rho(\bm{r})}{1+d_1\rho(\bm{r})}
\end{align}
with $w_1, d_1>0$. The second term in the denominator sets the upper limit
of $D$ as $|D|< w_1/d_1$ at extremely high density regions.
We adopt $d_1 = 1.0$ fm$^3$ and $w_1=742$ MeVfm$^8$ 
as prescribed in Ref.~\cite{Nakada15a}.

Once the HFB equation is solved, we can evaluate various observables.
The single-quasi-particle orbit includes the fractional occupation
of the single-particle (sp) orbit,
indicating the mixing of the sp orbits.
  It is useful to evaluate
the occupation probability $n_k$ and sp radius $r_k$
for $k$ orbital as
\begin{align}
n_k &= 4\pi \int \! dr\, r^2 \left|V_k(r)\right|^2 ,\\
r_k &= \sqrt{\frac{4\pi}{n_k} \int\! dr\, r^4 \left|V_k(r)\right|^2 }.
\end{align}

The second-order moment of the nuclear charge distribution 
is given by~\cite{Kurasawa19}
\begin{eqnarray}
&& r^2_{\rm ch} = \langle r^2 \rangle_{p} + r^2_p + \left( r^2_+ - r^2_-\right) \frac{N}{Z} \notag\\
&& \phantom{r^2_{\rm ch} =} + \langle r^2 \rangle_{W_p} + \langle r^2 \rangle_{W_n} \frac{N}{Z} + \delta\langle r^2\rangle_c \,,
\end{eqnarray}
where $r_p= 0.81$ fm, $r^2_\pm=(0.9)^2 \mp 0.06$ fm$^2$. The point neutron/proton density and spin-orbit density are given by
\begin{eqnarray}
&& \langle r^2 \rangle_q = \frac{4\pi}{N_q} \int \!dr \, r^4 \rho_q(r) \,,\\
&& \langle r^2 \rangle_{W_q} = - \frac{\mu_q\left( \hbar c\right)^2}{2\left(M c^2\right)^2} \frac{4\pi}{N_q} \int \!dr \, r^4 \nabla\cdot\vJ_q(r) \,,
\end{eqnarray}
where $q=n$ or $p$, namely, $N_n=N$, $N_p=Z$,
$\mu_n=1.793$, $\mu_p=-1.913$, and $M$ is the nucleon mass.
The last term is the relativistic correction:
\begin{align}
\delta\langle r^2\rangle_c = \frac{3 \left( \hbar c\right)^2}{4\left(M c^2\right)^2} + \frac{1}{2\mu_p} \langle r^2 \rangle_{W_p} \,.
\end{align}

\section{Adjustment of parameters}
\label{adjpara.sec}

\begin{figure*}[htb]
 \begin{center}
   \epsfig{file=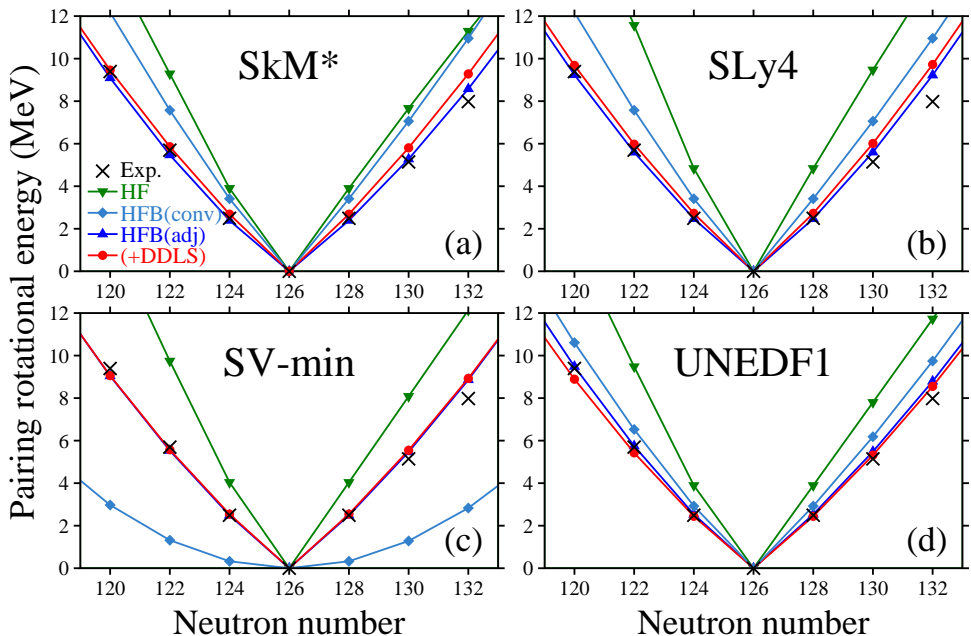, scale=0.35}
   \caption{
     Neutron pairing-rotational energies measured
     from a reference state $^{208}$Pb for
(a) SkM$^\ast$, (b) SLy4, (c) SV-min, and (d) UNEDF1 interactions. 
     HFB(conv) and HFB(adj) respectively indicate the HFB calculations with
  the conventional and present parametrizations of the pairing strength.
  See text for details.
       Experimental (black) values are evaluated from
       the experimental binding energies taken from Ref.~\cite{NNDC}
       by using Eq.~(\ref{pairing-rotational-energy}). }
    \label{pairrot.fig}
  \end{center}
\end{figure*}

\begin{figure*}[ht]
 \begin{center}
   \epsfig{file=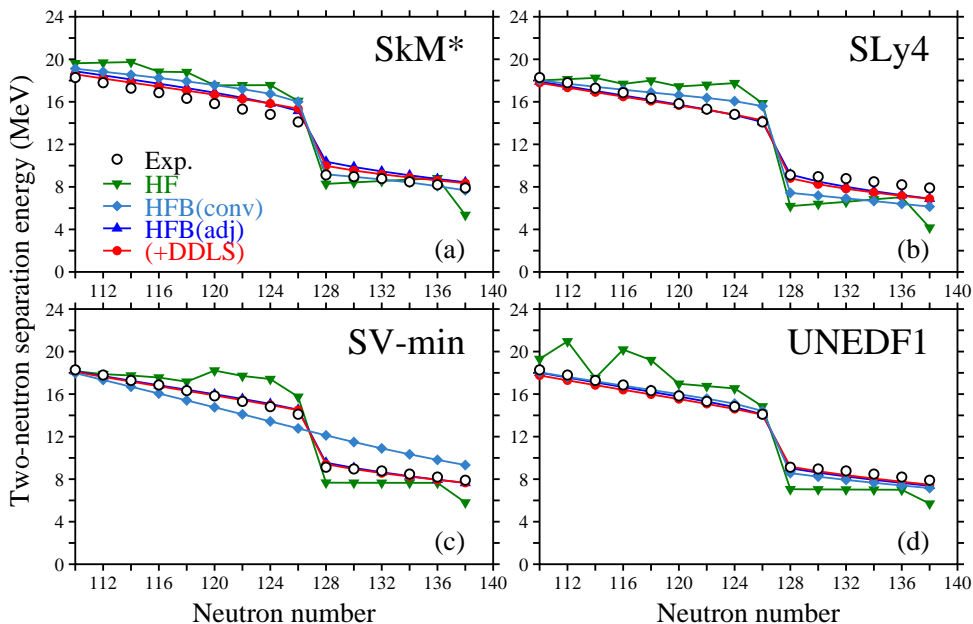, scale=0.35}
   \caption{Two-neutron separation energies of Pb isotopes
for (a) SkM$^\ast$, (b) SLy4, (c) SV-min, and (d) UNEDF1 interactions.
Experimental data are taken from \cite{NNDC}. 
  HFB(conv) and HFB(adj) respectively indicate the HFB calculations with
  the conventional and present parametrizations of the pairing strength.
See text for details.}
    \label{S2nPb.fig}
  \end{center}
\end{figure*}

The pairing interaction plays a vital role to determine
the configuration mixing near the Fermi level.
The pairing strength has been conventionally
determined in such a way that
theoretically obtained pairing gaps in even-even nuclei 
reproduce experimental odd-even mass differences.
However, no direct relationship between the calculated pairing gap 
and the experimental odd-even mass exists,
and there are a variety of definitions of 
theoretical pairing gaps and extracting the odd-even mass difference
from experiment~\cite{Satula98,Duguet01,Bertsch09}.
Recently, Ref.~\cite{Hinohara16}
proposed a new sound procedure to determine the
pairing strength using the pairing-rotational energy.
The pairing-rotational energy 
is defined by the second-order term of the ground-state energy
expanded in the neutron number $\Delta N=N-N_0$ as
\begin{align}
E(N,Z_0) = E(N_0,Z_0) - \lambda_n(N_0,Z_0) \Delta N
 +\frac{\left(\Delta N\right)^2}{2\mathfrak{I}(N_0,Z_0)},
\label{pairing-rotational-energy}
\end{align}
where $E(N_0,Z_0)$ denotes the ground-state energy 
of a nucleus with neutron number $N_0$ and proton number $Z_0$,
$\lambda_n(N_0,Z_0)= \left. dE/dN\right|_{N=N_0}\equiv
[E(N_0-2,Z_0)-E(N_0+2,Z_0)]/4$ is the chemical potential.
The last term of Eq.~(\ref{pairing-rotational-energy})
is the pairing-rotational energy, which 
can be evaluated only by using the experimental data for even-even nuclei.
In the present study,
the pairing strength $V_0$ of Eq.~(\ref{mixed.eq}) is determined
so as to reproduce the pairing-rotational energy around $^{208}$Pb
($N_0=126$ and $Z_0=82$) for each Skyrme functional.

Next, we explain how to implement the DDLS interaction
in the Skyrme functionals [See Eq.~(\ref{DDLSimplementation})].
We will search the two parameters $f_{ls}$ and $g_{ls}$ to reproduce
the isotope shift of the charge radius of Pb isotopes at $N>126$. 
In this procedure, we pay special attention
to the neutron sp level ordering around Fermi level in $^{208}$Pb.
It is known that neutron sp levels
around Fermi level ($2p_{1/2}$) are 
$0i_{13/2}$, $2p_{3/2}$, $1f_{5/2}$, $2p_{1/2}$, $1g_{9/2}$, $0i_{11/2}$,
and $0j_{15/2}$ in order~\cite{Gales78,Martin91,Martin93}.
As was pointed out in Ref.~\cite{Nakada15a}, if the $0i_{11/2}$ orbit
is located below the $1g_{9/2}$ orbit, 
the charge radius increases at $N>126$ drastically.
However, it contradicts the ordering of the sp levels,
which should be preserved.
Thus, we implement the DDLS and tune the parameters $f_{ls}$ and $g_{ls}$
under the condition that the neutron sp level ordering
is consistent with experimental data.
We do not pay much attention to the ordering
of three higher nodal orbitals, $2p_{3/2}$, $1f_{5/2}$, and $2p_{1/2}$
as they are almost degenerate and have a small impact on
changes in the isotope shift. 
After we find optimal $f_{ls}$ and $g_{ls}$ values, 
we re-optimize the pairing strengths $V_0$ by
following the procedure explained in the previous paragraph.
Hereafter we examine the following three types of calculations:
HF (without the pairing and DDLS interactions),
HFB (with the pairing interaction and without the DDLS interaction),
  and HFB+DDLS (with the pairing and DDLS interactions).

Figure~\ref{pairrot.fig} displays 
the pairing-rotational energies measured 
from a reference state $^{208}$Pb as a function of the neutron number
for the SkM$^\ast$, SLy4, SV-min, and UNEDF1 functionals.
We see that without the pairing correlations
the HF calculations do not reproduce the pairing-rotational
energy around $^{208}$Pb deduced from the experimental data.
With the optimized pairing and DDLS parameters,
the HFB(adj) and HFB+DDLS calculations nicely reproduce
the experimental data for all the Skyrme interactions employed in this paper.
To show further improvement with this parameter set,
Fig.~\ref{S2nPb.fig} plots two-neutron separation energies obtained
by HF, HFB, and HFB+DDLS calculations. 
The HFB(adj) and HFB+DDLS results are greatly improved compared to the results of HF and correctly describe the experimental data.

For the sake of comparison,
  we also show the results with the conventional parametrization
  for the pairing interaction,
  in which a fixed pairing strength is used for nuclei covering the whole
  nuclear chart [HFB(conv)].
For SkM$^\ast$ and SLy4, the pairing strengths are
 tuned to give a mean neutron gap of 1.245 MeV
 in $^{120}$Sn~\cite{Bennaceur05}.
The pairing strength of UNEDF1 is determined together with other
Skyrme parameters to reproduce nuclear matter properties and nuclear
masses~\cite{UNEDF1}.
For SV-min, the odd-even staggering is employed to determine the
pairing strength~\cite{SVmin}.
We see that the reproducibility of the experimental pairing rotational
energy and two-neutron separation energies are
not well with the conventional pairing strengths.
Describing the pairing rotational energies is important
to determine the pairing strength.
For the convenience of the reader,
we list all the optimized values of the pairing strength $V_0$,
$f_{ls}$ and $g_{ls}$ in the DDLS interaction in Table~\ref{DDLSpara.tab}.
The following discussions will be made based on
the results obtained with these parameter sets.

\begin{table}[ht]
  \caption{Pairing strength $V_0$ and the parameters of the DDLS interaction,
    $f_{ls}$ and $g_{ls}$, adopted in the present work.}
\begin{center}
  \begin{tabular}{cccccccccc}
    \hline\hline
  &&\multicolumn{3}{c}{HFB}&&\multicolumn{3}{c}{HFB+DDLS}\\    
    \cline{2-5}\cline{7-9}
    && $V_0$ (MeV)& $f_{ls}$&$g_{ls}$&&$V_0$ (MeV)& $f_{ls}$&$g_{ls}$ \\
\hline
  SkM$^\ast$&&$-$269 & 1.00 & 0.00&&$-$269 & 0.75 & 0.15 \\
  SLy4     && $-$321 & 1.00 & 0.00&&$-$323 & 0.80 & 0.20 \\
  SV-min   && $-$243 & 1.00 & 0.00&&$-$246 & 0.80 & 0.20 \\
  UNEDF1   && $-$248 & 1.00 & 0.00&&$-$246 & 0.80 & 0.10 \\
  \hline\hline
\end{tabular}
\end{center}
\label{DDLSpara.tab}
\end{table}

\section{Results and discussion}
\label{results.sec}

\begin{figure*}[htb]
\begin{center}
  \epsfig{file=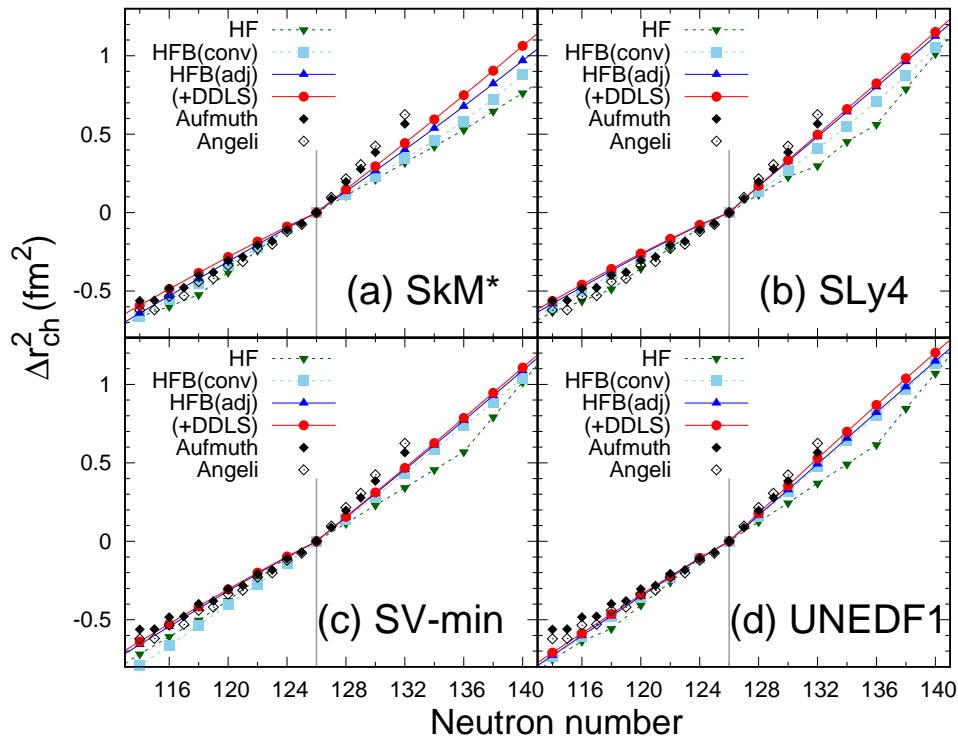, scale=1.2}                    
  \caption{Absolute difference of square charge radius of Pb isotopes
    from that of $^{208}$Pb as a function of the neutron number
    with HF, HFB, and HFB+DDLS calculations.
     HFB(conv) and HFB(adj) respectively indicate the HFB calculations with
  the conventional and present parametrizations of the pairing strength.
See text for details.
    The vertical lines indicate $N=126$.
    The (a) SkM$^\ast$, (b) SLy4, (c) SV-min, and (d) UNEDF1
    interactions are employed.
    Experimental data are taken from Refs.~\cite{Aufmuth87,Angeli13}.}
    \label{chargePb.fig}
  \end{center}
\end{figure*}

\subsection{Charge radius kink and pairing correlations}
\label{charge.sec}

Figure~\ref{chargePb.fig} plots
the difference of the square charge radii of Pb isotopes $r^2_{\rm ch}(N)$
from that of $^{208}$Pb defined by
\begin{align}
 \Delta r^2_{\rm ch}=r^2_{\rm ch}(N)-r^2_{\rm ch}(N_0)
\end{align}
with $N_0=126$.
Neither HF calculation adopted here reproduces
the kink structure at $N=126$, while a clear kink appears at $N=126$
when the proper pairing correlations are included [HFB(adj)].
 We also plot the results with the conventional parametrization
  for the pairing interaction [HFB(conv)], showing less reproducibility
  of the experimental data as was already pointed out in Ref.~\cite{Goriely15}. Hereafter we only discuss the results
with the present parametrization [HFB(adj)] as the HFB calculation.
In the present choice of the pairing strength,
the HFB already describes the kink behavior 
and is slightly improved by the HFB+DDLS calculation for 
all the Skyrme-type interactions adopted in this paper.
The agreement between the theory and experiment is satisfactory
as good as those obtained in Ref.~\cite{Nakada15a}.
We remark that in Ref.~\cite{Nakada15a} 
the DDLS effect appeared to be more drastic than
the present results because the different choices of the pairing 
and energy functionals.

\subsection{Pairing core swelling effect at $N>126$}
\label{swelling.sec}

We see that the large enhancement of the charge radius in Pb isotopes for $N>126$
is well reproduced by the inclusion of the adjusted pairing interaction.
This kink behavior is an indication of ``core'' swelling,
which is recently discussed for Ca isotopes for $N>28$
in the charge radius measurement~\cite{Ruiz16}
and interaction cross section measurement~\cite{Tanaka20}.
Here we show this sudden core swelling for Pb isotopes
can also be understood by considering the mechanism
recently proposed in Ref.~\cite{Horiuchi20} for spherical nuclei:
The occupation of nodal or $j$-lower ($j_<=l-1/2$ with $l>0$) orbits
can induce a sudden swelling of the ``core''
because the density distributions of those orbits have
a large overlap with the internal or core density~\cite{Horiuchi20}.
We remark that the configuration mixing regarding nuclear deformation was
discussed in Ref.~\cite{Horiuchi21a}.
In the present paper, we define the core nucleus as $^{208}$Pb
for $N>126$ and discuss the role of the valence neutrons
in this core swelling phenomenon.
We note that for $N>126$ the quantum number of the lowest orbit is $1g_{9/2}$.
However, as we see in Fig.~\ref{chargePb.fig},
the charge radius kink or the core swelling is not enough
in the HF calculation even though the nodal $1g_{9/2}$ orbit
is filling for $N=126$--$136$.

To understand the role of the pairing and DDLS interactions
in the core swelling phenomena in detail,
we list in Table~\ref{Pb.tab} 
the occupation numbers of the $1g_{9/2}$, $0i_{11/2}$, and $0j_{15/2}$ 
orbits and various root-mean-square (rms) radii
of $^{214}$Pb obtained by the HF, HFB, and HFB+DDLS calculations.
In the HF calculation, all the ``valence'' neutrons occupy
 the lowest $1g_{9/2}$ level as $n(1g_{9/2})=6$.
By including the pairing correlations, 
the occupation number is shared with these three orbits,
$1g_{9/2}$, $0i_{11/2}$, and $0j_{15/2}$.
The sum of these occupation numbers $n_v=n(1g_{9/2})+n(0i_{11/2})+n(0j_{15/2})$
is close to 6. Therefore, these three orbits can be regarded as the ``valence''
neutron orbits for $^{214}$Pb.

Here, we see that the pairing correlations
significantly change the rms sp radii of $1g_{9/2}$ orbit
and induce the mixing of the $1g_{9/2}$ and $0i_{11/2}$, and $0j_{15/2}$ orbits.
References~\cite{Reinhard94,Nakada15a} pointed out that
the mixing of the $0h_{11/2}$ orbit is a key to the understanding
of the kink behavior at $N>126$.
Without the pairing correlations,  
the $1g_{9/2}$ orbits still have a large rms sp radius.
The pairing correlations play a role to reduce 
the size of the $1g_{9/2}$ orbit
and induce the mixing of the $0i_{11/2}$ orbit.
We note that the $0i_{11/2}$ orbit have a smaller rms radius
than the $1g_{9/2}$ orbit.
Both the effects facilitate the core swelling because an overlap
between the valence and core densities becomes large as
in the same mechanism proposed in Ref.~\cite{Horiuchi20}.
The pairing correlations also induce the mixing of
the $0j_{15/2}$ orbit but its contribution to the core swelling
is expected to be not large. Because the orbit is nodeless and $j$-upper,
and its rms sp radius is the largest
among the three main orbits of the valence neutrons,
the probability of finding these neutrons is
distributed mostly near the nuclear surface.

We discuss the role of the DDLS interaction.
As expected from the results in Ref.~\cite{Nakada15a},
the DDLS plays a role to increase (decrease)
the radii of $j_<$ ($j_>=l+1/2$) orbits.
In fact, the rms sp radius of the $0i_{11/2}$ is enhanced
by the DDLS interaction. At the same time, the DDLS interaction also induces 
a little change of the configuration mixing
of the $1g_{9/2}$, $0i_{11/2}$, and $0j_{15/2}$ orbits.
In the present cases, the DDLS mainly enhances the rms sp radius of 
the $0i_{11/2}$ orbit, e.g., for SkM$^\ast$,
the enhancement is about 0.05 fm from that of the HFB result,
while the radius of the $0i_{13/2}$ orbit is almost unchanged by 0.005 fm
as the orbit belongs to the ``core''.
Consequently, its radius becomes closer to that of the $0i_{13/2}$, 6.42 fm,
resulting in a larger overlap between the core and valence nucleon densities
and hence the core swells.

\begin{table*}[htb]
\begin{center}
  \caption{Occupation numbers and rms sp neutron radii, in units of fm,
    of $^{214}$Pb near the Fermi level
    and these for the valence neutrons ($n_v$ and $r_v$).
    See text for details.
    Charge ($r_{\rm ch}$), rms point-proton ($r_p$),
    neutron ($r_n$), and matter  ($r_m$) radii
    are also listed in units of fm.}  
\label{Pb.tab}
\begin{tabular}{ccccccccccccccccc}
  \hline\hline
          &&    \multicolumn{3}{c}{SkM$^\ast$}&&\multicolumn{3}{c}{SLy4}&&\multicolumn{3}{c}{SV-min} && \multicolumn{3}{c}{UNEDF1}  \\
\cline{3-5}\cline{7-9}\cline{11-13}\cline{15-17}
 && HF& HFB & (+DDLS)&& HF& HFB & (+DDLS) && HF& HFB & (+DDLS)&& HF& HFB & (+DDLS)\\
\hline
$n(1g_{9/2})$  && 6&3.68&3.47 && 6 &3.13&3.02&&6&3.29&3.46&&6&3.09&2.70\\ 
$n(0i_{11/2})$ &&--& 0.76  &1.30&&--&1.56&1.78&&--&1.50&1.41&&--&1.51&2.16\\ 
$n(0j_{15/2})$ &&--& 1.40 & 0.97&&--&1.10&0.94&&--&0.91&0.83&&--&1.15&0.88\\
$n_v$         && 6& 5.84& 5.74&&6  &5.80&5.74&&--&5.71&5.74&&6&5.76&5.74\\
\hline
$r(1g_{9/2})$  &&6.57&6.47& 6.48&&6.71& 6.56&6.57&&6.49&6.37&6.38&&6.53&6.41&6.40\\    
$r(0i_{11/2})$ &&--   &6.31& 6.36&&--&6.39&6.41&&--&6.27&6.29&&--&6.25&6.28\\    
$r(0j_{15/2})$ &&--   &6.68& 6.68&&--&6.74&6.73&&--&6.58&6.61&&--&6.61&6.61\\    
$r_v$       &&6.57 &6.50 &6.49&&6.71&6.55&6.55&&6.49&6.38&6.39&&6.53&6.41&6.39\\
\hline
$r_{\rm ch}$&&5.52& 5.54& 5.55&&5.53&5.55&5.55&&5.52&5.53&5.53&&5.54&5.55&5.56\\
$r_p$&&5.48& 5.49& 5.50&&5.49&5.50&5.50&&5.48&5.49&5.48&&5.49&5.50&5.51\\   
$r_n$&&5.69& 5.69& 5.70&&5.69&5.70&5.70&&5.69&5.69&5.68&&5.70&5.70&5.71\\   
$r_m$&&5.61& 5.62& 5.63&&5.61&5.62&5.62&&5.61&5.61&5.60&&5.62&5.63&5.63\\   
\hline\hline
\end{tabular}
\end{center}
\end{table*}

We also calculate the average rms sp radius of the valence neutrons defined by
\begin{align}
  r^2_{v}=\frac{\sum_{\bar{n},\bar{l},\bar{j}\in v}n(\bar{n}\bar{l}_{\bar{j}})r^2(\bar{n}\bar{l}_{\bar{j}})}{\sum_{\bar{n},\bar{l},\bar{j}\in v}n(\bar{n}\bar{l}_{\bar{j}})}.
\end{align}
Here $\bar{n}\bar{l}\bar{j}$ runs over $1g_{9/2}$, $0i_{11/2}$, and $0j_{15/2}$.
The results are listed in Table~\ref{Pb.tab}.
Note that the $r_v$ value of the HF calculation
corresponds to the rms sp radius of $1g_{9/2}$ orbit.
The pairing interaction reduces the
radius of $1g_{9/2}$ and increases the mixing probability 
of the $0i_{11/2}$, which has a smaller sp radius compared to the others.
The $0j_{15/2}$ orbit has the largest radius among the others 
but the occupation number is small.
In total, the rms radius of the valence neutrons
is significantly reduced by the pairing interaction.
An interesting observation is that the proton and neutron rms radii
are enhanced when the pairing correlations are included
despite the rms radius of the valence neutrons being reduced.
The pairing correlation induces the core swelling.
We remark that in Ref.~\cite{Perera21} argued that
the pairing correlation reduces the charge radius kink at $N>126$
based on the covariant density functional theory.
Because in the covariant density functional theory
the $0i_{11/2}$ energy is lower than the $1g_{9/2}$ one,
the occupation number of the $0i_{11/2}$ orbit is reduced when
the pairing interaction is included, which
is in contrast to the present Skyrme density functional cases.
We find that the DDLS interaction plays a role in reducing the rms radius 
of the valence neutrons for all the Skyrme interactions
and further enhances the rms proton radius. 
The reduction of the valence neutron radius 
is essential to induce the core swelling.

\begin{figure*}[ht]
\begin{center}
  \epsfig{file=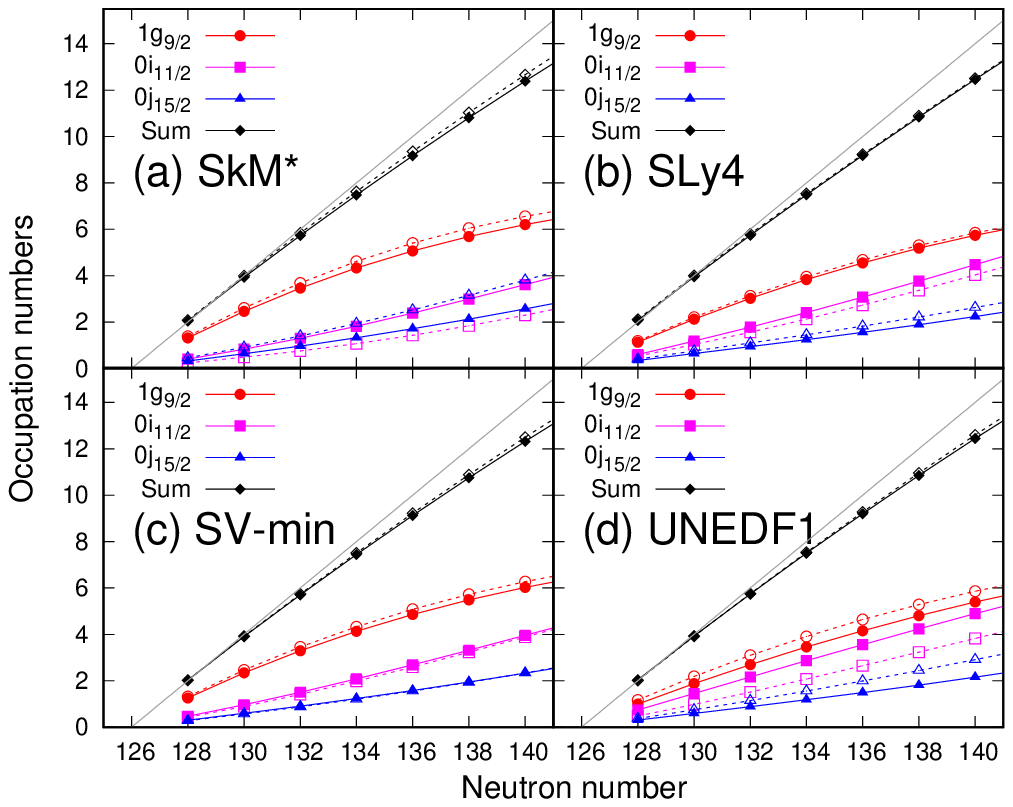, scale=1.2}                    
  \caption{Occupation numbers of the $1g_{9/2}$, $0i_{11/2}$,
    and $0j_{15/2}$ orbits of Pb isotopes calculated
    with HFB (open symbols with dashed lines)
    and HFB+DDLS (closed symbols with solid lines).
    The sum of these occupation numbers are also drawn.    
    The line indicate $N-126$ line, which corresponds to
    the maximum occupation numbers in the assumption of $N=126$
    closed configuration for the $^{208}$Pb core.
    The (a) SkM$^\ast$, (b) SLy4, (c) SV-min, and (d) UNEDF1
    interactions are employed.}
    \label{occPb.fig}
  \end{center}
\end{figure*}

Figure~\ref{occPb.fig} plots the occupation numbers
of the $1g_{9/2}$, $0i_{11/2}$, and $0j_{15/2}$
orbits, which give the three largest occupations 
for the Pb isotopes at $N>126$, as a function of the neutron number.
We also show the occupation numbers of the sum of these three orbits.
Since these sums almost follow $N-126$ with $126<N<140$,
these three orbits can still be regarded as the valence neutrons
as the contributions from higher-lying orbits, i.e.,
the $1g_{7/2}$, $2d_{5/2}$, $3s_{1/2}$, and $2d_{3/2}$ orbits,
are still not large in this mass region, at most $\approx 1/14$
of the valence neutrons at $N=140$.
By including the DDLS interaction, as expected from Ref.~\cite{Nakada15a},
a relatively large enhancement of the occupation of $0i_{11/2}$ orbit
is obtained for SkM$^\ast$ and UNEDF1, while
the occupation number of the $0j_{15/2}$ orbit is reduced.
Since all the orbits other than the $0j_{15/2}$ orbit
can contribute to swelling the core~\cite{Horiuchi20},
further core swelling is expected when the DDLS interaction is implemented.
In fact, the rms valence neutron orbits reduces and
the rms proton radius is enhanced by the inclusion of the DDLS interaction.
We note that in Fig.~\ref{chargePb.fig},  
the improvement of the charge radius results was not significant for
the SLy4 and SV-min interaction.
Because the pairing correlations already incorporate
possible reduction of the valence neutron orbits,
there is no room to enhance the proton radius with the inclusion
of the DDLS interaction.
In fact, virtually no change is obtained in the occupation numbers 
with the HFB and HFB+DDLS results for SLy4 and SV-min
as seen in Fig.~\ref{occPb.fig}.

For larger $N$ regions for $N>140$, in addition to the three main orbits
the occupation numbers of
the $1g_{7/2}$, $2d_{5/2}$, $3s_{1/2}$, and $2d_{3/2}$ orbits becomes large.
They are nodal orbits that can have a large overlap
with the sp density in the internal regions.
Further development of the core swelling is expected for $N>140$.

\subsection{Effects of the pairing correlations to other observables}
\label{diff.sec}

\begin{figure*}[ht]
\begin{center}
  \epsfig{file=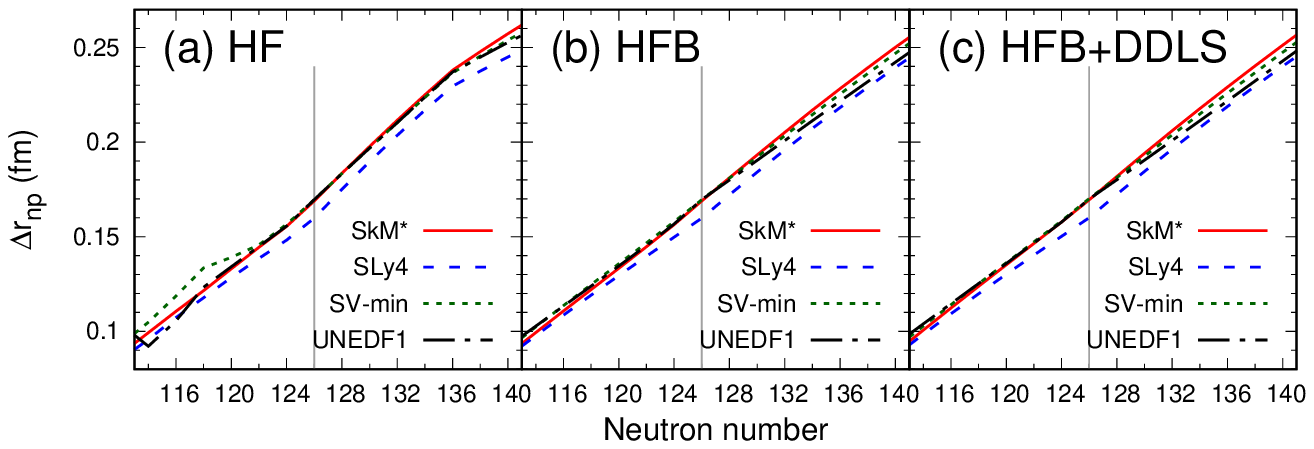, scale=1.2}                    
  \caption{Neutron skin thickness of Pb isotopes
    as a function of the neutron number calculated with
  (a) HF, (b) HFB, and (c) HFB+DDLS.}
    \label{skinPb.fig}
  \end{center}
\end{figure*}

It is also interesting to note that the matter radius increases
even though the valence neutron orbits shrink
by the pairing interaction. In this last subsection,
we discuss how this structure change affects
the density profile or the nuclear size properties.
Figure~\ref{skinPb.fig} plots the neutron skin thickness,
defined by $\Delta r_{np}=r_n-r_p$, of Pb isotopes at $114<N<140$.
In the HF calculation [Fig.~\ref{skinPb.fig} (a)],
some wiggles are found due to changes
of the outermost neutron orbits, e.g., from $1g_{9/2}$ to $0i_{11/2}$ at $N=136$.
In the HFB and HFB+DDLS calculations respectively drawn in
Figs.~\ref{skinPb.fig} (b) and (c), we do not see that ill behavior, showing
the neutron skin thickness is a robust quantity.
The neutron number dependence of the skin thickness exhibits
good linearity with small interaction dependence,
if the pairing interaction is properly chosen.

The HFB and HFB+DDLS calculations induce the mixing
of the main three orbits $1g_{9/2}$, $0i_{11/2}$, and $0j_{15/2}$ for $N>126$.
The occupation probabilities of the sp orbits near the Fermi level
strongly affect the density distribution near the nuclear surface.
Determining such density profiles is useful to study
various nuclear structure as exemplified
in Refs.~\cite{Hatakeyama18, Choudhary20, Choudhary21}.
To probe the degree of the configuration mixing,
we show the diffuseness parameter
of the matter density distributions, which can accurately be deduced
from proton-elastic scattering~\cite{Hatakeyama18}.
Suppose that the nuclear density is approximated as the two-parameter Fermi
distribution (2pF)
\begin{align}
\rho_{\rm 2pF}(r)=\frac{\rho_c}{1+\exp{[(r-R)/a]}},
\end{align}
where $\rho_c$ is determined from
the normalization condition $4\pi\int_0^\infty dr\, r^2 \rho_{\rm 2pF}(r)=A$,
for given the radius ($R$) and diffuseness ($a$) parameters.
As prescribed in Ref.~\cite{Hatakeyama18},
the $R$ and $a$ values are fixed in such a way 
to minimize the rms deviation of the density distributions obtained
by the present calculations.
The nuclear diffuseness evolves depending on
the characteristics of the sp wave function
of the valence neutrons~\cite{Horiuchi21b}:
Nodal (nodeless) orbits near the Fermi level always enhance (reduce)
the nuclear surface diffuseness.

\begin{figure*}[ht]
\begin{center}
  \epsfig{file=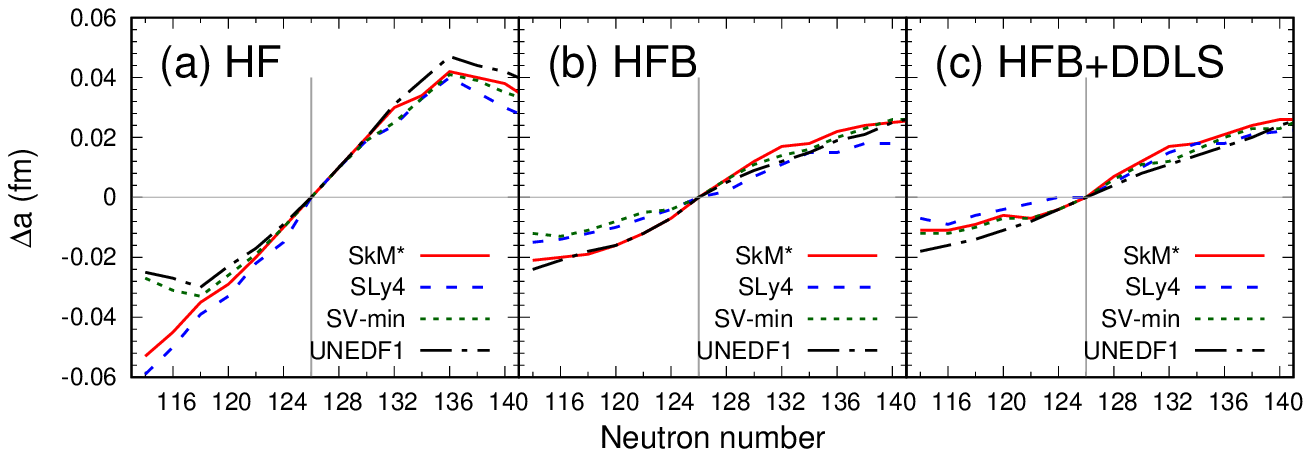, scale=1.2}                    
  \caption{Absolute difference of the nuclear surface diffuseness
    measured from $^{208}$Pb calculated with
  (a) HF, (b) HFB, and (c) HFB+DDLS.}
    \label{diffPb.fig}
  \end{center}
\end{figure*}

Figure~\ref{diffPb.fig} displays the absolute difference of the
nuclear surface diffuseness of $N>126$ from that of $^{208}$Pb,
$\Delta a(N)=a(N)-a(N_0=126)$.
The $\Delta a$ becomes largest in the HF case
at $N=136$ and reduces rapidly for $N>136$.
This is simply because
the $1g_{9/2}$ orbit plays a role in enhancing the nuclear diffuseness
at $126<N<136$ and the occupation of the $0i_{11/2}$ orbit reduces the
nuclear diffuseness at $N>136$.
With the pairing interaction, the HFB case,
the enhancement of the nuclear diffuseness is not significant
compared to the HF case due to
the occupation of ``sharp'' $0i_{11/2}$ and $0j_{15/2}$ orbits
in addition to the occupation of ``diffused'' $1g_{9/2}$ orbit.

\begin{figure}[ht]
\begin{center}
  \epsfig{file=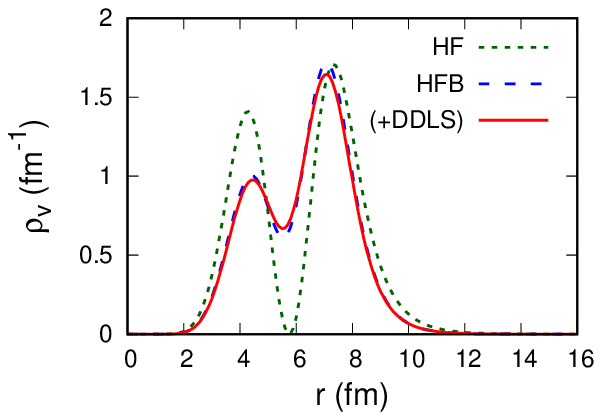, scale=1.3}                    
  \caption{Valence neutron densities of $^{214}$Pb with the SkM$^\ast$ interaction. See text for details.}
    \label{spdens.fig}
  \end{center}
\end{figure}

To see changes in the densities near the nuclear surface,
Figure~\ref{spdens.fig} compares the sp densities
of the valence neutrons,
defined by $\rho_v(r)=4\pi r^2\sum_{i\in v}|V_i(r)|^2$,
of $^{214}$Pb with the SkM$^\ast$ interaction
for the HF, HFB, and HFB+DDLS calculations.
Obviously, a clear two-peak structure is found for the HF
calculation originating in the $(1g_{9/2})^2$ configuration.
As we already discussed, a shrinkage of the $1g_{9/2}$ orbit
and the mixing of $0i_{11/2}$ and $0j_{15/2}$ orbits occur
with the pairing interaction.
Since the nodeless high angular momentum orbits, i.e.,
the $0i_{11/2}$ and $0j_{15/2}$ orbits, are mixed,
the nuclear surface diffuseness is reduced~\cite{Horiuchi21b}.
A systematic measurement of the nuclear surface
diffuseness will reveal the degree of the configuration mixing of Pb isotopes
for $N>126$, which is closely related to the core swelling phenomenon.

\section{Conclusion}
\label{conclusion.sec}

We have revisited the kink behavior of the isotope shift of
the charge radius of Pb isotopes at $N>126$
in terms of the Skyrme-Hartree-Fock-Bogoliubov (HFB) theory.
Four standard Skyrme-density functionals, SkM$^\ast$, SLy4,
SV-min, and UNEDF1 have been employed.
The strengths of the pairing interaction
for each Skyrme functional has been determined so as 
to reproduce the pairing-rotational energy around $^{208}$Pb
and the parameters of
the density-dependent spin-orbit (DDLS) term has been fixed
for the first time for the Skyrme-density functionals.

With the present choice of the pairing interaction,
the kink behavior of the isotope shift of the charge radius
can be reproduced fairly well, while no kink appears without
the pairing interaction.
The DDLS interaction further increases the charge radius
 but its effect is small compared to the pairing correlations.
In Pb isotopes at $N>126$, 
the pairing correlations act to induce the mixing
of the $0i_{11/2}$ orbit and the shrinkage of the $1g_{9/2}$ orbits.
Since the $0i_{11/2}$ orbit is smaller than the $1g_{9/2}$ orbit,
in total the ``valence'' neutron orbits shrink
and ``core'' swelling occurs as interpreted in Ref.~\cite{Horiuchi20}.
This novel pairing core swelling effect is imprinted in
the isotope shifts of the charge radius at $N>126$
as well as the density profile near the nuclear surface exhibiting
a moderate change of the nuclear surface diffuseness. 
It is interesting to explore the possible mechanism of the charge radius kink
in different mass regions as it reflects 
configurations near the Fermi level, which was recently realized in
Sn isotopes at $N>132$~\cite{Gorges19}.

\acknowledgments
This work was in part supported by JSPS KAKENHI Grants No.\ 18K03635.
We acknowledge the collaborative research program 2021, 
Information Initiative Center, Hokkaido University.

\end{document}